\documentclass{article}

\usepackage[pdftex]{graphicx}  
\usepackage{lineno,hyperref}
\modulolinenumbers[5]

\usepackage{gensymb}
\usepackage{amsmath}
\usepackage{caption}
\usepackage{subcaption}
\usepackage{authblk}

\title{Improving count rate capability of timing RPCs by increasing the detector working temperature}
\author[1]{A. Blanco}
\author[1,2]{P. Fonte}
\author[1]{L. Lopes}
\author[1]{J. Saraiva}

\affil[1]{Laboratory of Instrumentation and Experimental Particles Physics, Coimbra, Portugal}
\affil[2]{Coimbra Polytechnic - ISEC, Coimbra, Portugal}

\begin{document}
	\maketitle
\begin{abstract}
This communication describes test beam results, focusing on detection efficiency and timing precision, of common float glass timing Resistive Plate Chambers (tRPCs) exposed to a $2.7$~GeV proton beam and operated at above ambient temperature in order to increase the count rate capability of the chambers, by exploiting the reduction in the resistivity of the glass electrodes. Results suggest that the count rate capability can be extended at least up to $1500~Hz/cm^2$ when the detector is operated at $40.6^{\circ}$C without noticeable loss of efficiency or timing precision degradation with values of $90\%$ and $100$~ps, respectively, for this specific timing RPC arrangement.  
\end{abstract}

\section{Introduction}
tRPC have traditionally been used with a relatively low particle flux ($<<$ kHz/cm$^2$) due to the inherent limitation to the counting rate imposed by the commonly used float glass electrode resistivity. Since tRPCs are one of the main large-area timing detectors, extension of its counting rate capability is of great interest for future High Energy Particle (HEP) experiments, were the luminosity is expected to increase considerably.

Attempts have already been made to increase the count rate capability by using materials with lower electrical resistivity compared to the commonly used float glass, such as ceramics \cite{LOPES20074, NAUMANN2011138}, special glasses \cite{WANG2010151} or some technical plastics \cite{PlasticsImad}. As a result, the operation of small area detectors was successfully achieved, but the implementation of the medium/large area detectors failed due to the lack of homogeneity of the materials, which present low electrical resistivity paths, resulting in an unstable behavior of the detector. Another possibility, still very little explored, is to decrease the resistivity of standard float glass by increasing the operational temperature of the detectors, providing a ten-fold decrease in resistivity every $25^{\circ}$C \cite{GONZALEZDIAZ200572}.

This communication describes test beam result, focusing on detection efficiency and timing precision, of common float glass tRPCs exposed up to incident particle fluxes up to $1500~Hz/cm^2$ and operated up to $40.6^{\circ}$C.

\section{Experimental setup}
\label{sec:setup}
\begin{figure}
\centering 
\includegraphics[width=\linewidth]{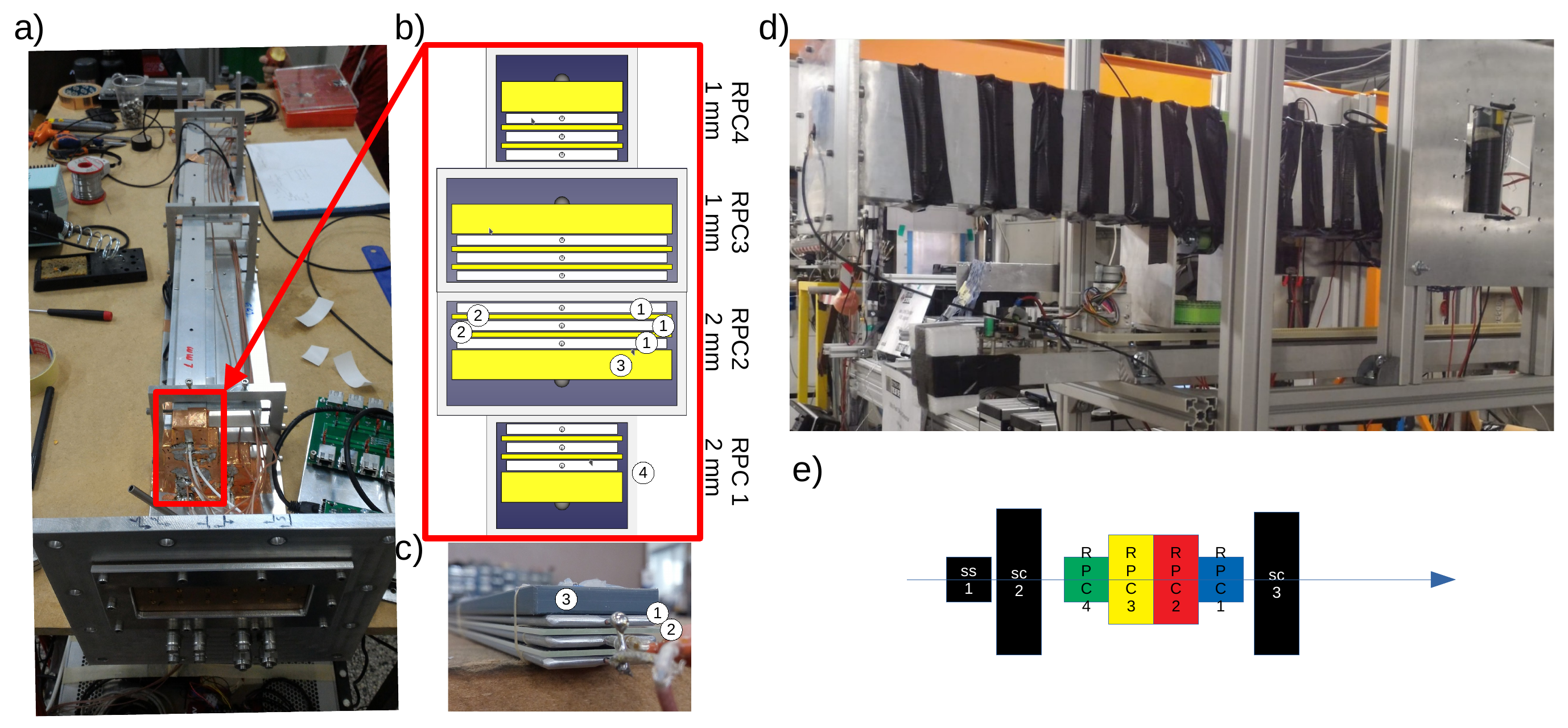}
\caption{a). Internal arrangement of the tRPC. b) Cross section of the four chambers arrangement and c) close-up photograph of one of the $22$~mm wide chamber, with: 1- Al electrodes, 2- glass electrodes, 3- PVC pressure plate and 4- 2 mm thick aluminum shielding tube. d) Panoramic view of the setup in the beam line, showing the tRPC gas box, surrounded by the heating wire, and the last scintillator of the telescope. e) Cross section of the scintillator and tRPC in the beam line.}
\label{fig:setup}
\end{figure}

The experimental arrangement consist of four individually shielded strip-like tRPC stacked one on top of the other, see figures \ref{fig:setup}.a and \ref{fig:setup}.b. Each individual chamber consists of three aluminum ($2$~mm thickness) and two glass (soda-lime) electrodes with a length of $750$~mm and two different widths $22$~mm ($RPC_1$ and $RPC_4$) and $44$~mm ($RPC_2$ and $RPC_3$). Two of the chambers ($RPC_3$ and $RPC_4$) are equipped with $1$~mm glass, (bulk resistivity of $\approx 4$x$10^{12}$~$\Omega$cm at $25$~$^\circ$C) while the others ($RPC_2$ and $RPC_1$) are equipped with $2$~mm thick glass (bulk resistivity of $\approx 10$x$10^{12}$~$\Omega$cm at $25$~$^\circ$C).  All of the electrodes have rounded edges and the glass exceeds the aluminum in size by $1$~mm (in both dimensions) to prevent discharges, see figures \ref{fig:setup}.b and \ref{fig:setup}.c. The gap is defined by PEEK (Polyetheretherketone) mono-filaments of $0.270$~mm diameter, spaced approximately by $100$~mm along the chamber. The ensemble is housed inside individual aluminum tubes (shielding) and compressed by springs on top of each mono-filament that apply a controlled force through a PVC (Polyvinyl chloride) plate that distributes the force.

High-voltage (HV) close to $6$~kV is applied to the central aluminum electrode via $1$~M$\Omega$ resistors and high voltage cables, while the outer electrodes are grounded and the glass electrodes are kept electrically floating. Insulation to the shielding tube walls is assured by a triple-layer KAPTON\texttrademark\hspace{0.1cm} adhesive laminate. An end-shield made of aluminium foil is glued to both ends of the aluminium tube to produce a fully shielded element. The signals are collected, at both ends of each chamber, by coaxial cables through $2$~nF $HV$ coupling capacitors and extracted from the gas box through the front panel via RF MMCX connectors.  

The gas box is equipped with temperatures sensor and a heating element, capable to increase the working temperature homogeneously in the inner chambers up to at least $40.6^{\circ}$C.
A set of fast plastic scintillators define a beam line (telescope). The telescope is composed of three paralalepipeds $80$~x~$30$~x~$20$ mm$^{3}$ scintillators (Bicron BC420), SC$_1$, SC$_2$ and SC$_3$  placed as it is sketched in figure \ref{fig:setup}.e), read out each of them by two fast PMTs (Hamamatsu H6533). The $20$~mm side faces the beam passing through $30$~mm scintillator for maximum timing precision.

Both tRPC and the PMTs signals are feed to fast Front End Electronics (FEE) \cite{HADES_FEE} (for PMT readout the amplification stage was removed) capable of measuring time and charge in a single channel. The resulting signals are read out by the TRB board \cite{TRB3} equipped with $128$ multi-hit TDC (TDC-in-FPGA technology) channels with a time precision better than $20$~ps.
 
The chambers were operated in open gas loop with a mixture of $97$\%  $C_{2} H_{2} F_{4}$ and $3$\% $SF_{6}$, exposed to $2.7$~GeV proton beam, with a typical diameter of $40$~mm, fluxes up to $1500$~Hz/cm$^{2}$ and working temperatures up to $40.6^{\circ}$C. Since the gain of an tRPC depends on the pressure and temperature of the gas, the applied $HV$ is modulated as a function of these two variables according to \cite{Lopes_2014} in order to keep the gain constant. 

\section{Methods}
The test focused on obtaining the timing precision, $\sigma_{RPC_i}$, and efficiency, $\epsilon_i$, of each individual tRPC ${i=1...4}$, as a function of the particle flux, from a few $Hz/cm^2$ up to $1500$~$Hz/cm^2$ (measured using the scintillator telescope) for different temperatures, from $21^{\circ}$C up to $40.5^{\circ}$C. 
For each particle detected in the tRPC ${i}$, the following variables are computed: time as $T_i = (T_{il} + T_{ir})/2$, where $T_{il}$ and $T_{ir}$ are the left and right times and charge as $Q_i = (Q_{il} + Q_{ir})/2$, with $Q_{il}$ and $Q_{ir}$ the left and right charges. Any time differences involving $T_i$ are corrected as a function of $Q_i$ using the typical slewing correction. Similar procedure is followed for the scintillator signals. tRPC timing precision, $\sigma_{RPC_i}$, is calculated by performing the time differences among tRPC and the two front cintillators, $SC_1$ and $SC_2$, in the telescope and resolving the following system of equations:
\begin{equation}
\begin{pmatrix}
1 & 1 & 0\\
1 & 0 & 1\\
0 & 1 & 1\\
\end{pmatrix}
\begin{pmatrix}
\sigma_{RPC_i}^2\\
\sigma_{SC_1}^2\\
\sigma_{SC_2}^2\\
\end{pmatrix}
=
\begin{pmatrix}
\sigma_{\Delta_{RPC_i-SC_1}}^2\\
\sigma_{\Delta_{RPC_i-SC_2}}^2\\
\sigma_{\Delta_{SC1-SC_2}}^2\\
\end{pmatrix}
\end{equation}
where $\sigma_{RPC_i}$, $\sigma_{SC_1}$ and $\sigma_{SC_2}$ are the time precision of RPC {i}, $SC_1$ and $SC_2$ and $\sigma_{\Delta_{RPC_i-SC_1}}$, $\sigma_{\Delta_{RPC_i-SC_2}}$ and $\sigma_{\Delta_{SC_1-SC_2}}$ are the sigma of time difference distribution of $RPC_i-SC_1$, $RPC_i-SC_2$ and $SC_1-SC_2$. Typical calculated timing precision for the scintilators is around $35$~ps sigma. Efficiency, $\epsilon_i$, of each individual tRPC, is calculated as the ratio of the number of events with signal in both ends of a given tRPC ${i}$ and the number of events with signal in the three scintillators in the telescope. The $SC_1$ scintillator has a vertical width of $20$~mm, while the $RPC_1$ and $RPC_4$ have $22$~mm, therefore a misalignment between the tRPCs and the scintillator can create geometric inefficiencies, as will be shown below. In addition, due to the small size of the telescope, the efficiency obtained should be understood as a lower limit.

\section{Results}
\begin{figure}[h]
\centering 
\includegraphics[width=0.85\linewidth]{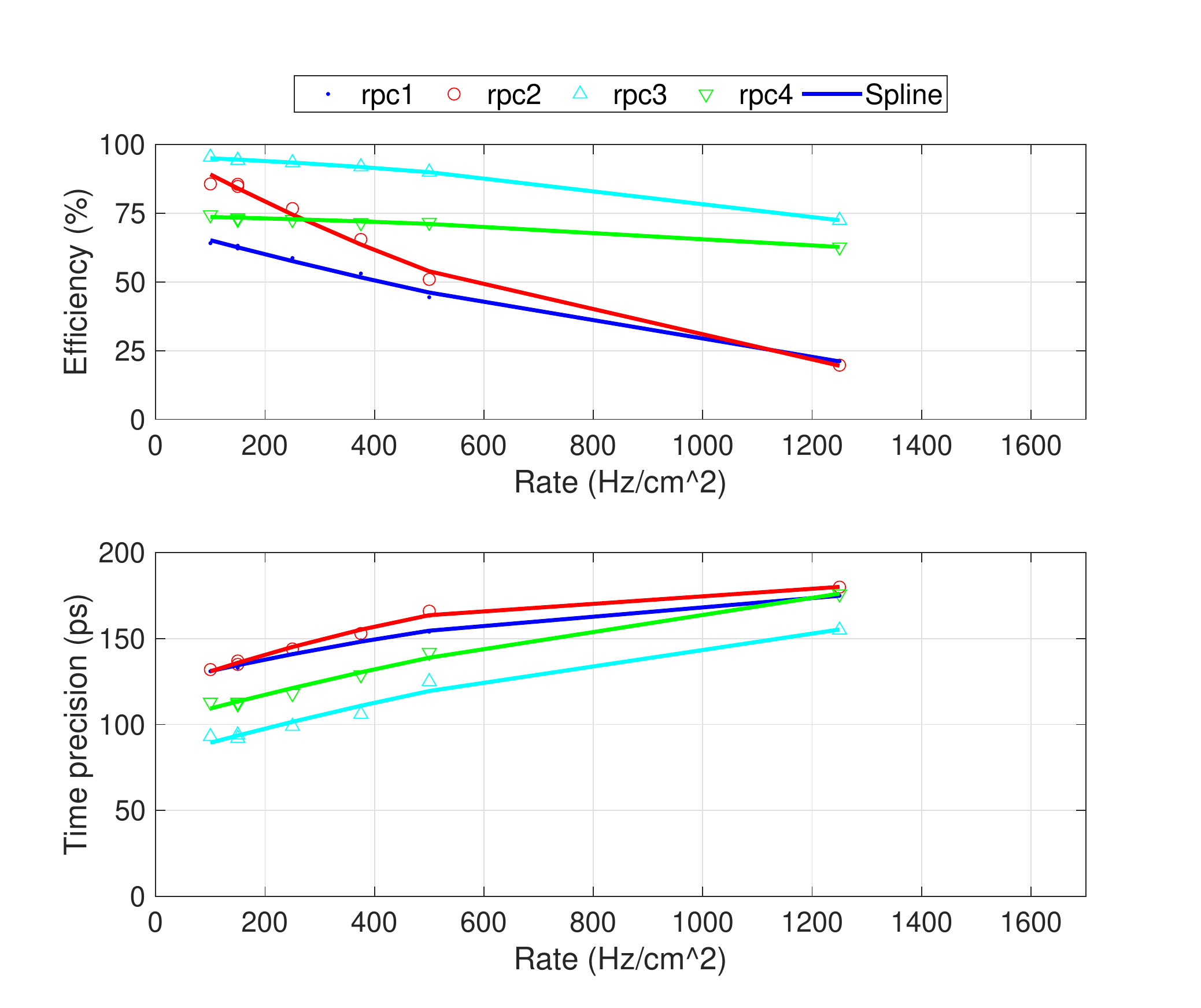}
\caption{a) Efficiency and b) timing precision of the tRPC as a function of the particle flux for a reduced electric field of $443$~Td and a working temperature of $21^{\circ}$C. A spline has been added to guide the eyes.}
\label{fig:eff_TvsRate}
\end{figure}

Figures \ref{fig:eff_TvsRate}.a and \ref{fig:eff_TvsRate}.b show the efficiency and timing precision as a function of the incident particle flux density at a working temperature of $21^{\circ}$C and reduced electric field of $443$~Td. Both figures show the loss of efficiency and deterioration of timing precision as the incident particle flux increases from $100$~Hz/cm$^{2}$ up to $1500$~Hz/cm$^{2}$. Apart from the already mentioned difference in efficiency (due to geometrical reasons) between the $22$~mm and $44$~mm tRPCs, it is observed that the tRPCs with 2 mm glass ($RPC_2$ and $RPC_1$) lose efficiency much faster and have a worse timing precision. This difference is due to two factors. On the one hand the thickness of the glass and on the other hand the resistivity itself which is $2$-$3$ times lower in the $1$ mm glass, giving in combination a factor of $4$-$6$ in resistance. 

\begin{figure}[h]
	\centering 
	\includegraphics[width=0.85\linewidth]{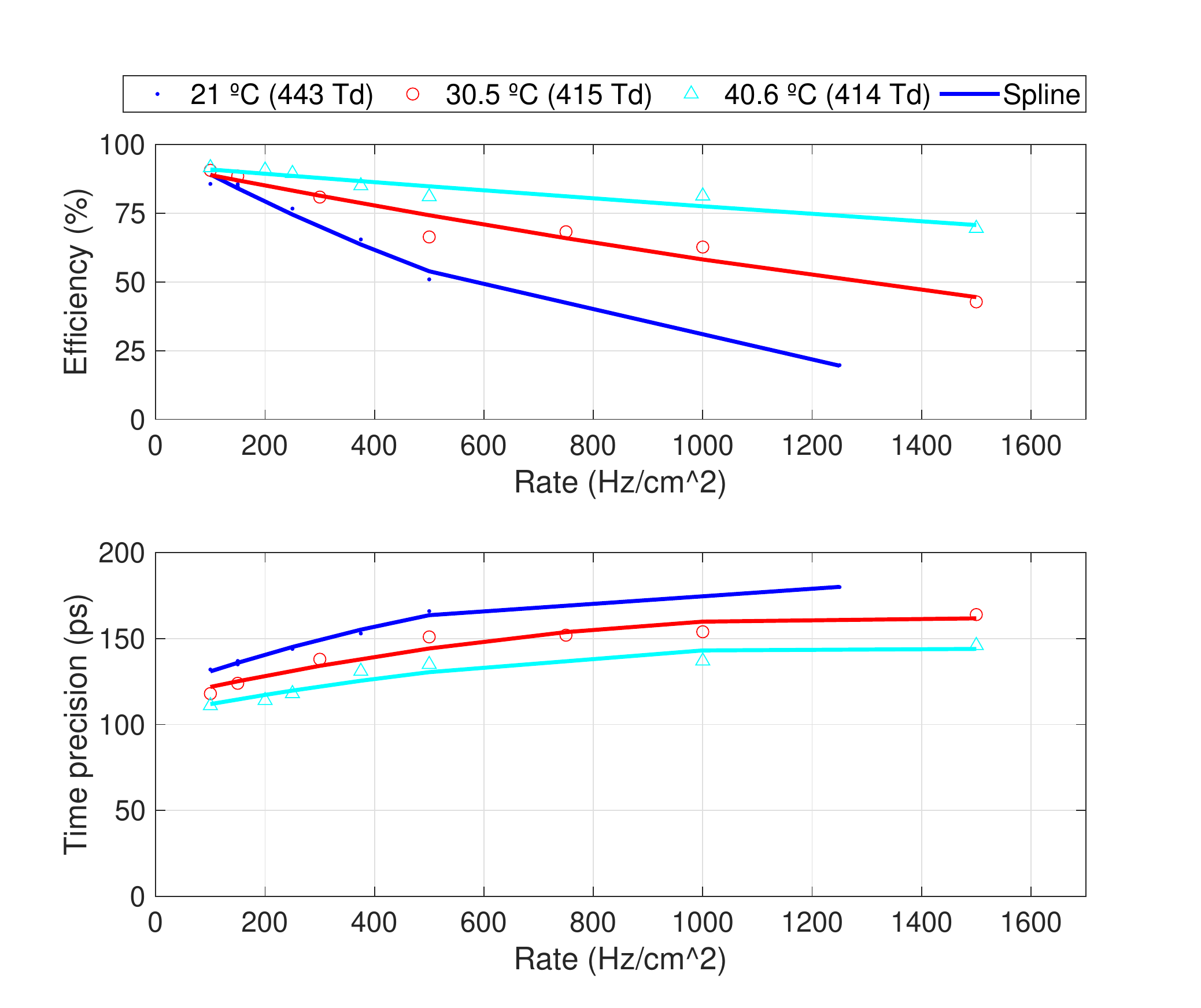}
	\caption{a) Efficiency and b) timing precision of the tRPC with glass thicknesses of $2$~mm as a function of particle flux for three different working temperatures: $21^{\circ}$C, $30.5^{\circ}$C and $40.6^{\circ}$C. A spline has been added to guide the eyes.}
	\label{fig:eff_TvsTemp_2mm}
\end{figure}

\begin{figure}
\centering 
\includegraphics[width=0.85\linewidth]{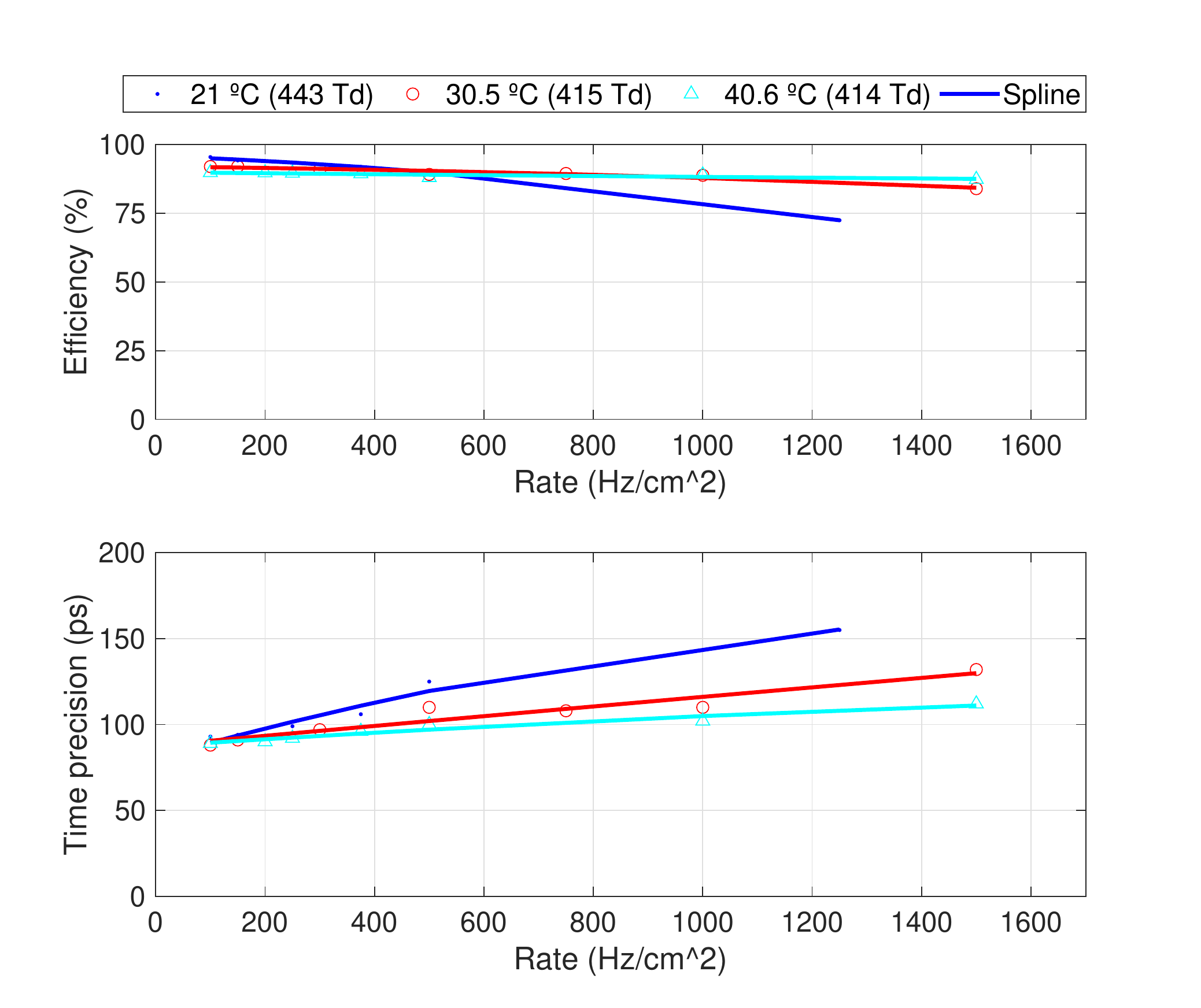}
\caption{a) Efficiency and b) timing precision of the tRPC with glass thicknesses of $1$~mm as a function of particle flux for three different working temperatures: $21^{\circ}$C, $30.5^{\circ}$C and $40.6^{\circ}$C. A spline has been added to guide the eyes.}
\label{fig:eff_TvsTemp_1mm}
\end{figure}

Figures \ref{fig:eff_TvsTemp_2mm}.a and \ref{fig:eff_TvsTemp_1mm}.a show the efficiency of the tRPC with $2$~mm and $1$~mm thick glass respectively as a function of the incident particle flux from  $100$~Hz/cm$^{2}$ up to $1500$~Hz/cm$^{2}$ for three different working temperatures $21^{\circ}$C, $30.5^{\circ}$C and $40.6^{\circ}$C. The efficiency recovery with increasing operating temperature (due to decreasing resistivity) is evident being much higher for the $1$~mm chambers (due to lower resistance) becoming basically independent of the incident particle flux (at least up to $1500$~Hz/cm$^{2}$) for a temperature of $40.6^{\circ}$C.  Chambers built with $2$~mm glass show only a modest recovery. Figures \ref{fig:eff_TvsTemp_2mm}.b and \ref{fig:eff_TvsTemp_1mm}.b shows the time precision for the same conditions mentioned for efficiency. Again, the recovery in timing precision is observed as the operating temperature increases, being more evident for the $1$~mm chambers, remaining at a level of approximately $100$~ps up to $1500$~Hz/cm$^{2}$.

\section{Conclusions}
In this work we have shown that increasing the working temperature of a tRPC can substantially improve its counting rate capability. In particular, individually shielded strip-like tRPC with an active area of $750$~x~$44$~mm equipped with $4$ gaps of $0.270$~um, show the same efficiency, close to $90\%$, and timing precision, close to $100$~ps, over a range of incident particle fluxes up to $1500$~Hz/cm$^{2}$ when their working temperature is raised to $40.6^{\circ}$C. This contrasts with a $20$ percentage points loss of efficiency and a worsening of temporal precision of more than $60$~ps when operated at $21^{\circ}$C. 

This way of improving the counting rate capability of a tRPC detector can be very interesting as it allows to extend its counting rate capability in a very simple way.

\section{Acknowledgments}
This work was supported by Fundação para a Ciência e Tecnologia, Portugal, in the framework of the project CERN/FIS-INS/0009/2019. 

\end{document}